\documentclass[aps,prl,reprint,twocolumn,twoside,showpacs,floatfix,letterpaper,superscriptaddress,bibnotes,footinbib]{revtex4-1}

\usepackage{graphicx}
\usepackage{dcolumn}
\usepackage{bm}
\usepackage{natbib}
\usepackage{multirow}
\usepackage{amsmath}
\usepackage{amssymb}
\usepackage{mathrsfs}
\usepackage{bbold}
\usepackage{hyperref}
\usepackage{aas_macros_prl}
\usepackage{color}

\newcommand{\hMpc}{h^{-1}{\rm Mpc}}
\newcommand{\hGpc}{h^{-1}{\rm Gpc}}

\newcommand{\void}{\mathrm{v}}

\newcommand{\rp}{\bar{r}_\mathrm{p}}
\newcommand{\rs}{r_s}
\newcommand{\dc}{\delta_c}

\begin{document}

\title{Universal Density Profile for Cosmic Voids}

\author{Nico Hamaus}
\email{hamaus@iap.fr}
\affiliation{Sorbonne Universit\'es, UPMC Univ Paris 06, UMR 7095, Institut d'Astrophysique de Paris, F-75014, Paris, France}
\affiliation{CNRS, UMR 7095, Institut d'Astrophysique de Paris, F-75014, Paris, France}

\author{P. M. Sutter}
\affiliation{Sorbonne Universit\'es, UPMC Univ Paris 06, UMR 7095, Institut d'Astrophysique de Paris, F-75014, Paris, France}
\affiliation{CNRS, UMR 7095, Institut d'Astrophysique de Paris, F-75014, Paris, France}
\affiliation{Center for Cosmology and Astroparticle Physics, Ohio State University, Columbus, Ohio 43210, USA}

\author{Benjamin D. Wandelt}
\affiliation{Sorbonne Universit\'es, UPMC Univ Paris 06, UMR 7095, Institut d'Astrophysique de Paris, F-75014, Paris, France}
\affiliation{CNRS, UMR 7095, Institut d'Astrophysique de Paris, F-75014, Paris, France}
\affiliation{Department of Physics, University of Illinois at Urbana-Champaign, Urbana, Illinois 61801, USA}


\begin{abstract}
We present a simple empirical function for the average density profile of cosmic voids, identified via the watershed technique in $\Lambda$CDM $N$-body simulations. This function is universal across void size and redshift, accurately describing a large radial range of scales around void centers with only two free parameters. In analogy to halo density profiles, these parameters describe the scale radius and the central density of voids. While we initially start with a more general four-parameter model, we find two of its parameters to be redundant, as they follow linear trends with the scale radius in two distinct regimes of the void sample, separated by its compensation scale. Assuming linear theory, we derive an analytic formula for the velocity profile of voids and find an excellent agreement with the numerical data as well. In our companion paper [Sutter \emph{et al.}, \mnras~\textbf{442}, 462 (2014)] the presented density profile is shown to be universal even across tracer type, properly describing voids defined in halo and galaxy distributions of varying sparsity, allowing us to relate various void populations by simple rescalings. This provides a powerful framework to match theory and simulations with observational data, opening up promising perspectives to constrain competing models of cosmology and gravity.
\end{abstract}


\maketitle

\textit{Introduction.}---While tremendous effort has been conducted studying the properties of dark matter halos, cosmic voids have largely been unappreciated by the broad scientific community. However, as voids occupy the most underdense regions in the Universe, and constitute the dominant volume fraction of it, they are promising independent probes to test our theories of structure formation and cosmology. For example, voids are the ideal laboratories for studies of dark energy (e.g., Refs.~\cite{Granett2008,Biswas2010,Lavaux2012,Bos2012,Cai2014a}) and modified gravity (e.g., Refs.~\cite{Li2011,Clampitt2013,Spolyar2013,Cai2014b}), as the importance of ordinary gravitating matter is mitigated in their interior. Unlike dark matter halos, voids are in addition more closely related to the initial conditions of the Universe, thanks to the limited number of phase-space foldings occurring inside of them~\cite{Shandarin2012,Falck2012,Neyrinck2012,Abel2012,Neyrinck2013,Leclercq2013}.

A fundamental quantity to describe the structure of voids in a statistical sense is their spherically averaged density profile. In contrast to the well-known formulas parametrizing density profiles of simulated dark matter halos (e.g., Refs.~\cite{Einasto1965,Burkert1995,Navarro1997,Moore1999}), rather few models for void density profiles have been developed, mainly focusing on the central regions~\cite{Colberg2005,Padilla2005,Lavaux2012,Ricciardelli2013,Ricciardelli2014}, rarely taking into account the compensation walls outside the void~\cite{Paz2013}. In this Letter we present a simple formula that is able to accurately describe the density profile around voids of any size and redshift, out to large distances from their center. Although we focus our attention on dark matter simulations here, our companion paper~\cite{Sutter2013a} extends the analysis to voids defined in other tracer types, such as dark matter halos and mock galaxies of various number densities, yielding consistent results. Thus, given the excellent agreement between voids found in mocks and in real observations~\cite{Sutter2012a,Sutter2013b,Ceccarelli2013,Paz2013,Pisani2013,Nadathur2014}, our results are relevant for observational data as well.

\textit{Simulations.}---We analyze outputs of the 2HOT $N$-body code~\cite{Warren2013} that evolved $2048^3$ cold dark matter particles in a $1\hGpc$ box of a Planck cosmology~\cite{Planck2013}. The snapshots are randomly subsampled to match a mean particle number density of $\bar{n}=2\times10^{-2}h^3\mathrm{Mpc}^{-3}$, corresponding to an average particle separation of $\rp\simeq3.7\hMpc$. This is comparable to the sampling density of modern galaxy surveys, such as \cite{SDSS, DES, EUCLID}. We then generate void catalogs using a modified version of the \textsc{zobov} code~\cite{Neyrinck2008,Lavaux2012,Sutter2012a}, which finds density minima in a Voronoi tessellation of the tracer particles and grows basins around them by applying the watershed transform~\cite{Platen2007}. This uncovers a nested hierarchy of voids and subvoids, all of which we include in our analysis. We restrict ourselves to zones with underdensity barrier $\delta\le-0.8$ when merging them into voids and define void centers as the mean of the void's particle positions, weighted by their Voronoi cell volume $V_\mathrm{c}$ (see Refs.~\cite{Lavaux2012,Sutter2012a}). Finally, an effective void radius $r_\void$ is defined as the radius of a sphere comprising the same volume as the watershed region that delimits the void.

\begin{figure*}[!t]
\centering
\resizebox{\hsize}{!}{
\includegraphics[trim=0 0 0 0,clip]{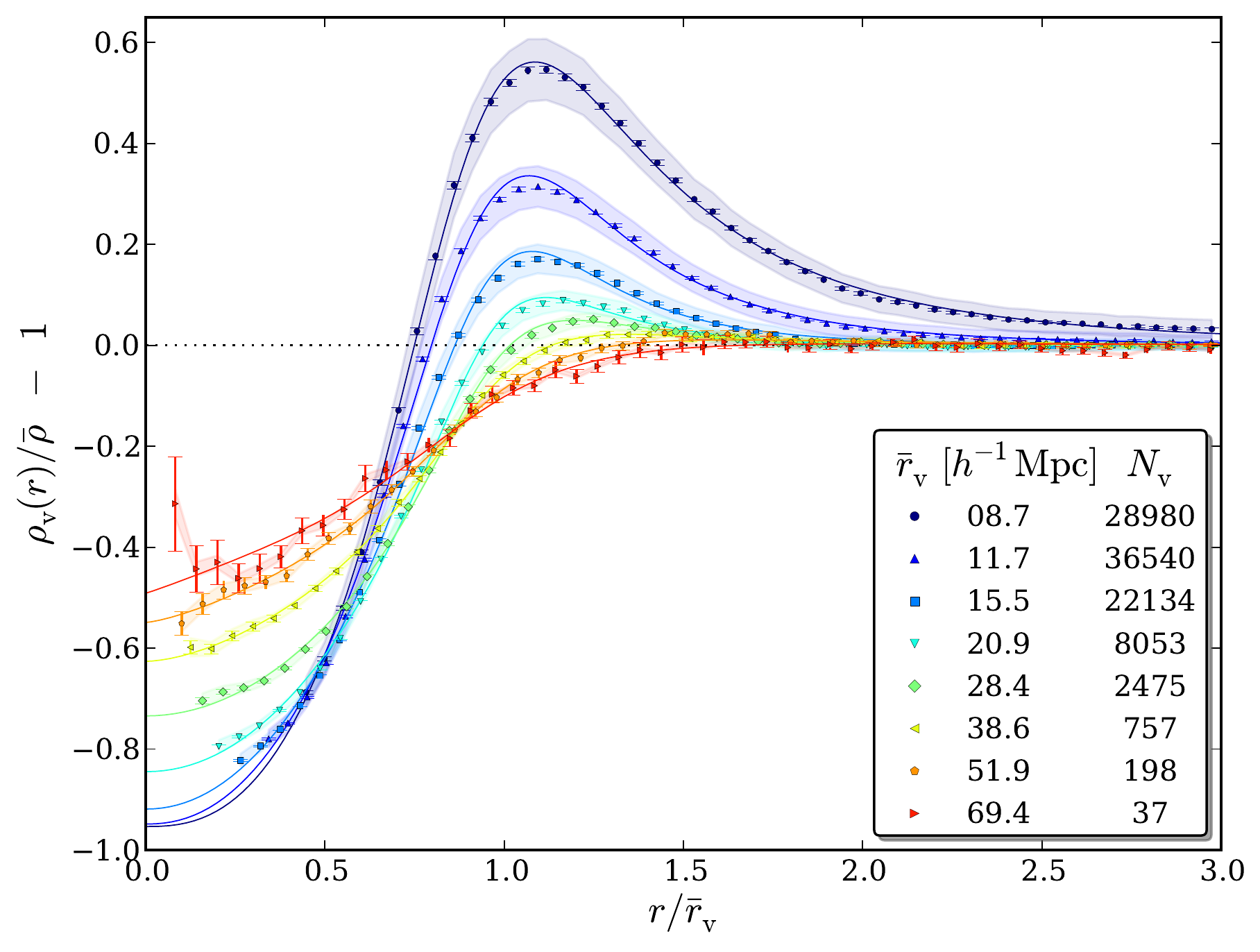}
\includegraphics[trim=0 0 0 0,clip]{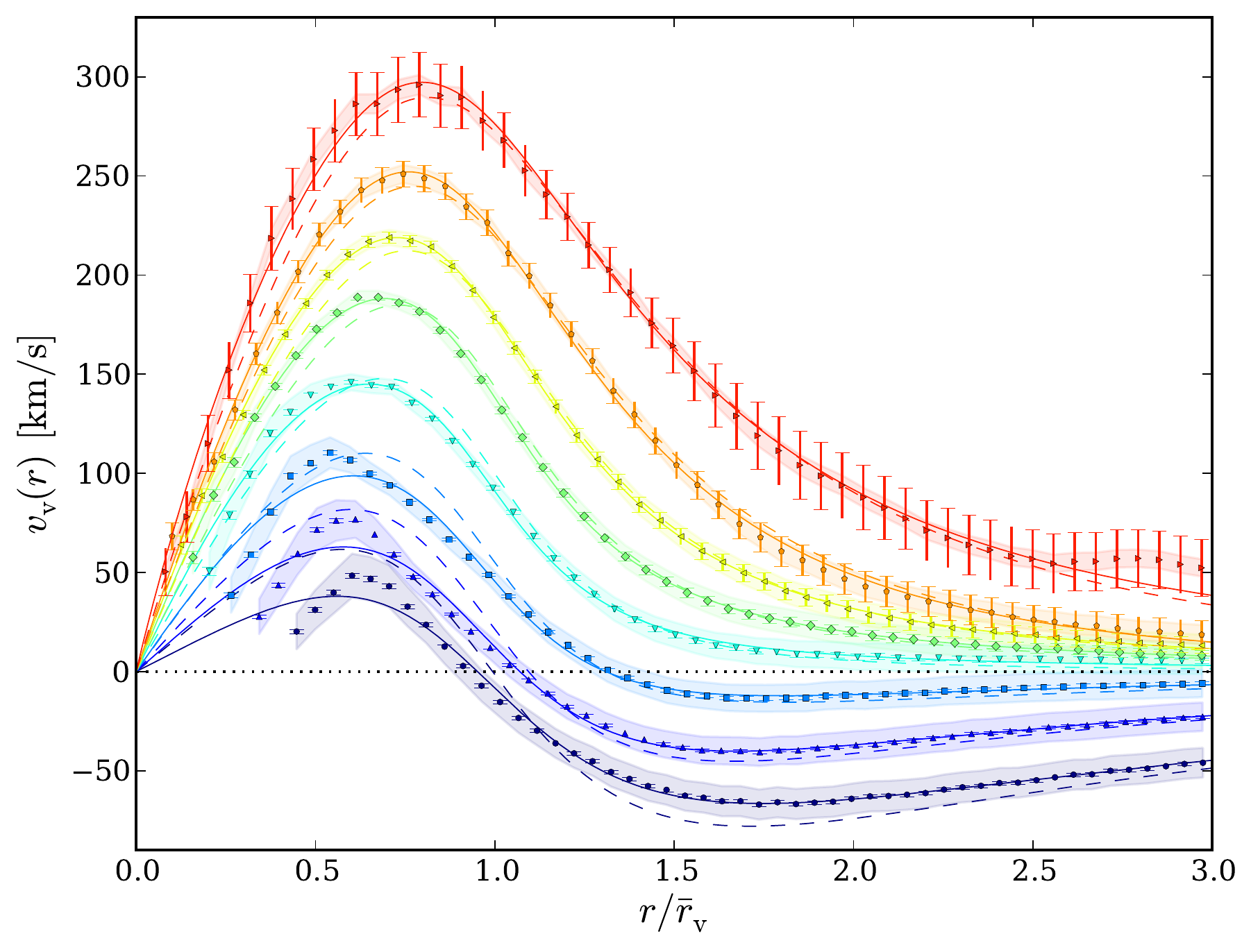}}
\caption{Stacked density (left) and velocity (right) profiles of voids at redshift zero in eight contiguous bins in void radius with mean values and void counts indicated in the inset. Shaded regions depict the standard deviation $\sigma$ within each of the stacks (scaled down by 20 for visibility), while error bars show standard errors on the mean profile $\sigma/\sqrt{N_\void}$. Solid lines represent our best-fit solutions from Eq.~(\ref{dprofile}) for density and from Eqs.~(\ref{vprofile}) and (\ref{intprofile}) for velocity profiles. Dashed lines show the linear theory predictions obtained from evaluating the velocity profile equation at the best-fit parameters obtained from the density stacks.}
\label{fig1}
\end{figure*}

\textit{Density profile.}---We define the void density profile as the spherically averaged relative deviation of mass density around a void center from the mean value $\bar{\rho}$ across the Universe, $\rho_\void(r)/\bar{\rho} - 1$. Using the tracer particles, the density in a radial shell of thickness $2\delta r$ at distance $r$ from a void center at the origin can be estimated as
\begin{equation}
 \rho_\void(r) = \frac{3}{4\pi}\sum_i\frac{m_i(\bm{r}_i)\Theta(r_i)}{(r+\delta r)^3-(r-\delta r)^3}\;, \label{rho}
\end{equation}
where $m_i$ is the mass of particle $i$, $\bm{r}_i$ its coordinate vector of length $r_i$, and $\Theta(r_i) \equiv \vartheta[r_i-(r-\delta r)]\vartheta[-r_i+(r+\delta r)]$ combines two Heaviside step functions $\vartheta$ to define the radial bin. In our simulations we use dark matter particles of equal mass and calculate the density profile of every void out to three times its effective radius $r_\void$. In order to avoid contamination from resolution effects, we include only voids with radii larger than twice the mean particle separation, $r_\void>2\rp$, and discard density estimates from Eq.~(\ref{rho}) at $r<\rp$. We then average (stack) all void profiles within eight contiguous logarithmic bins in void radius, to account for the poor statistics of the largest voids. The resulting stacks are shown with different symbols in the left-hand panel of Fig.~\ref{fig1}, where shaded regions depict the standard deviation $\sigma$ among all $N_\void$ voids within each stack, scaled down by a factor of 20 for visibility. Error bars show $\sigma/\sqrt{N_\void}$, the standard error on the mean profile.

As expected, stacked voids are deeply underdense inside, with their central density increasing with void size. In addition, the variance of underdense regions is suppressed compared to overdense ones~\cite{Bernardeau2013}, yielding the smallest error bars in the centers of the emptiest voids. However, note that the void-to-void scatter in the profile decreases towards the largest voids, as can be seen from the shaded regions in Fig.~\ref{fig1}. The profiles all exhibit overdense compensation walls~\cite{Bertschinger1985,White1990} with a maximum located slightly outside their effective void radius, shifting outwards for larger voids. The height of the compensation wall decreases with void size, causing the inner profile slope to become shallower and the wall to widen. This trend divides all voids into being either overcompensated or undercompensated, depending on whether the total mass within their compensation wall exceeds or falls behind their missing mass in the center, respectively~\cite{Hamaus2014}. Ultimately, at sufficiently large distances to the void center, all profiles approach the mean background density.

We propose a simple empirical formula that accurately captures the properties described above:
\begin{equation}
 \frac{\rho_\void(r)}{\bar{\rho}} - 1 = \dc\,\frac{1-(r/\rs)^\alpha}{1+(r/r_\void)^\beta}\;, \label{dprofile}
\end{equation}
where $\dc$ is the central density contrast, $\rs$ a scale radius at which $\rho_\void=\bar{\rho}$, and $\alpha$ and $\beta$ determine the inner and outer slope of the void's compensation wall, respectively. The best fits of this four-parameter model to the void density stacks are shown as solid lines in the left-hand panel of Fig.~\ref{fig1}. The concordance with the numerical data is exquisite everywhere.

\textit{Velocity profile.}---We estimate the velocity profile of tracer particles around void centers by calculating
\begin{equation}
 v_\void(r) = \frac{1}{N(r)}\sum_i\bm{v}_i(\bm{r}_i)\cdot\frac{\bm{r}_i}{r_i}\;V_\mathrm{c}(\bm{r}_i)\Theta(r_i)
\end{equation}
for every void and then averaging over all void radii in a given bin. Here, $\bm{v}_i$ is the particle velocity vector, $V_\mathrm{c}(\bm{r}_i)$ the Voronoi cell volume of a particle located at $\bm{r}_i$, and $N(r) \equiv \sum_iV_\mathrm{c}(\bm{r}_i)\Theta(r_i)$. Using the Voronoi volumes $V_\mathrm{c}$ as weights ensures a volumetric representation of the velocity field \footnote{\protect{G}.~Lavaux (private communication)}.

The right-hand panel of Fig.~\ref{fig1} depicts the resulting velocity stacks using the same void radius bins as for the density stacks. Note that a positive velocity implies outflow of tracer particles from the void center, while a negative one denotes infall. As the largest voids are undercompensated (void in void~\cite{Sheth2004}), i.e. the total mass in their surrounding does not make up for the missing mass in their interior, they are characterized by outflow in the entire distance range. Tracer velocities increase almost linearly from the void center until they reach a maximum located slightly below the effective void radius of each sample, which indicates the increasing influence of the overdense compensation wall. When passing the latter, tracer velocities are continuously decreasing again in amplitude and approach zero in the large distance limit.

Small voids may exhibit infall velocities~\cite{Cai2014a,Ceccarelli2013,Paz2013}, as they can be overcompensated (void in cloud~\cite{Sheth2004}). This causes a sign change in their velocity profile around the void's effective radius beyond which matter is flowing onto its compensation wall, ultimately leading to a collapse of the void. Moreover, because small voids are more underdense in the interior, their velocity profile is more nonlinear and less accurately sampled there. The distinction between overcompensation and undercompensation can directly be inferred from velocities, since only overcompensated voids feature a sign change in their velocity profile, while undercompensated ones do not. Consequently, the flow of tracer particles around precisely compensated voids vanishes already at a finite distance to the void center and remains zero outwards. By slightly shifting the void radius bins, we determined this to be the case for voids with $\bar{r}_\void\simeq17.6\hMpc$ in our sample, which we denote as the compensation scale. It can also be inferred via clustering analysis in Fourier space, as compensated structures do not generate any large-scale power~\cite{Hamaus2014}. We checked that the compensation scales obtained from these two independent methods agree very accurately, indicating a strong link between the spatial and dynamical characteristics of voids.

In linear theory the velocity profile can be related to the density using~\cite{Peebles1993}
\begin{equation}
 v_\void(r) = -\frac{1}{3}\Omega_\mathrm{m}^\gamma H r\Delta(r)\;, \label{vprofile}
\end{equation}
where $\Omega_\mathrm{m}$ is the relative matter content in the Universe, $\gamma\simeq0.55$ the growth index of matter perturbations, $H$ the Hubble constant, and $\Delta(r)$ the integrated density contrast defined as
\begin{equation}
 \Delta(r) = \frac{3}{r^3}\int_0^r\left(\frac{\rho_\void(q)}{\bar{\rho}}-1\right)q^2 \mathrm{d}q\;. \label{dcontrast}
\end{equation}
With Eq.~(\ref{dprofile}), this integral yields
\begin{multline}
 \Delta(r) = \dc\,{}_2F_1\!\!\left[1,\frac{3}{\beta},\frac{3}{\beta}+1,-(r/r_\void)^\beta\right]\\
 -\frac{3\dc(r/\rs)^\alpha}{\alpha+3}\,{}_2F_1\!\!\left[1,\frac{\alpha+3}{\beta},\frac{\alpha+3}{\beta}+1,-(r/r_\void)^\beta\right]\;, \label{intprofile}
\end{multline}
where ${}_2F_1$ is the Gauss hypergeometric function. When plugged into Eq.~(\ref{vprofile}), we can use this analytic formula to fit the velocity profiles obtained from our simulations; the results are shown as solid lines in the right-hand panel of Fig.~\ref{fig1}. As for the density profiles, the quality of the fits is remarkable, especially for large voids. Only the interiors of smaller voids show stronger discrepancy, which is mainly due to the decreasing validity of linear theory, i.e., Eq.~(\ref{vprofile}). We obtain best-fit parameter values that are very similar to the ones resulting from the density stacks above. In fact, evaluating the velocity profile at the best-fit parameters obtained from the density stacks yields almost identical results, as indicated by the dashed lines in Fig.~\ref{fig1}.

With the explicit form for the integrated density profile in Eq.~(\ref{intprofile}), it is straightforward to determine the void's uncompensated mass, defined as~\cite{Hamaus2014}
\begin{equation}
 \delta m = \lim_{r\to\infty}\frac{4\pi}{3}\bar{\rho}r^3\Delta(r)\;.
\end{equation}
The limit exists only for $\beta>\alpha+3$ and yields
\begin{equation}
 \delta m = \frac{4\pi^2\bar{\rho}r_\void^3\dc}{\beta}\left\{\csc\left(3\pi/\beta\right)-(r_\void/\rs)^\alpha\csc\left[(\alpha+3)\pi/\beta\right]\right\};
\end{equation}
i.e., compensated voids with $\delta m = 0$ satisfy the relation
\begin{equation}
 (\rs/r_\void)^\alpha = \frac{\sin(3\pi/\beta)}{\sin\left[(\alpha+3)\pi/\beta\right]}\;, \label{compensated}
\end{equation}
independently of $\dc$.

\begin{figure}[!t]
\centering
\resizebox{\hsize}{!}{
\includegraphics[trim=0 0 0 0,clip]{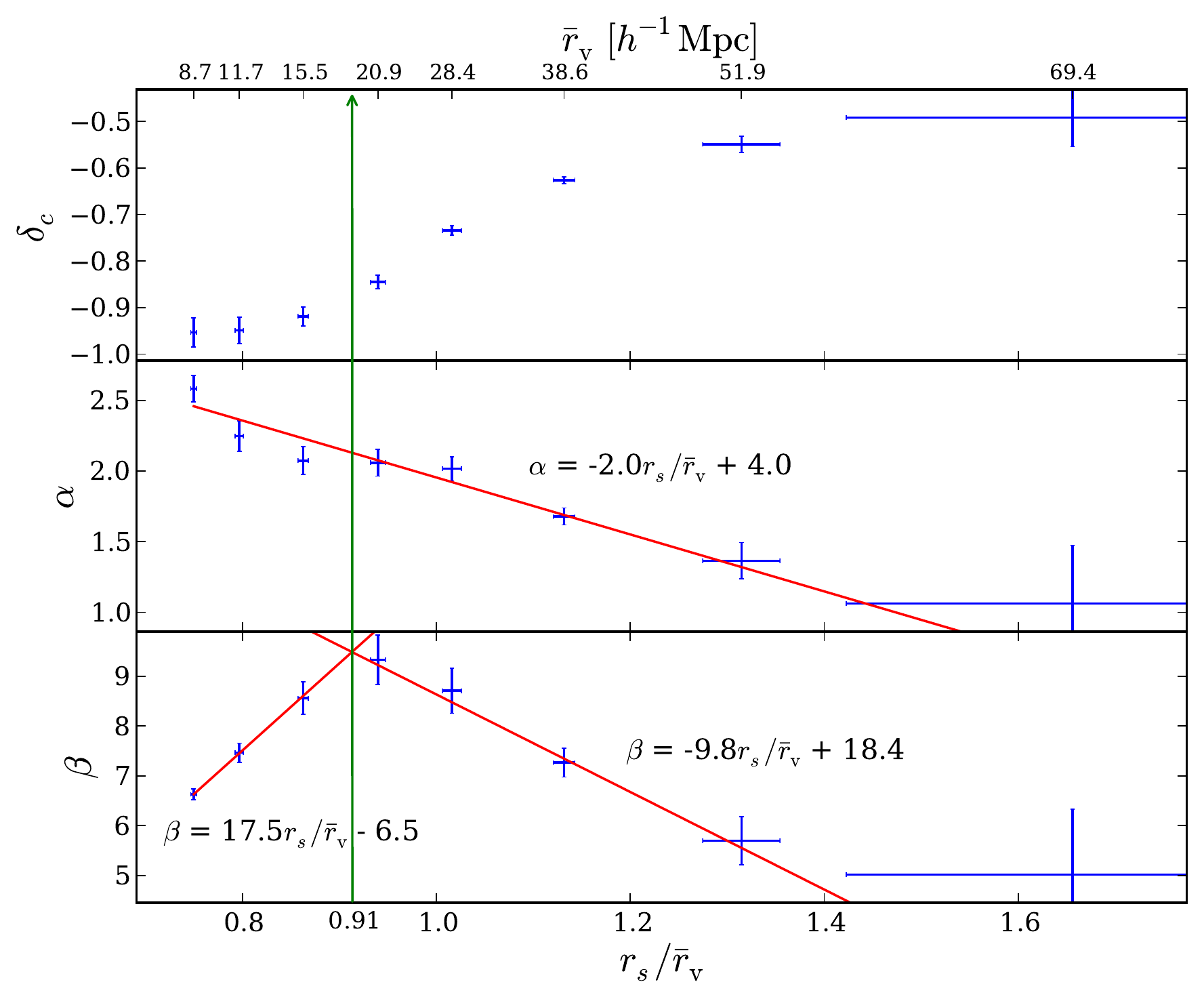}}
\caption{Best-fit parameters obtained from the density stacks of Fig.~\ref{fig1}. Error bars designate $95\%$ confidence regions and straight lines show linear regressions through the data points, with corresponding best-fit values stated alongside. The vertical line indicates the compensation scale of the void sample.}
\label{fig2}
\end{figure}

\begin{figure*}[!t]
\centering
\resizebox{\hsize}{!}{
\includegraphics[trim=0 0 0 0,clip]{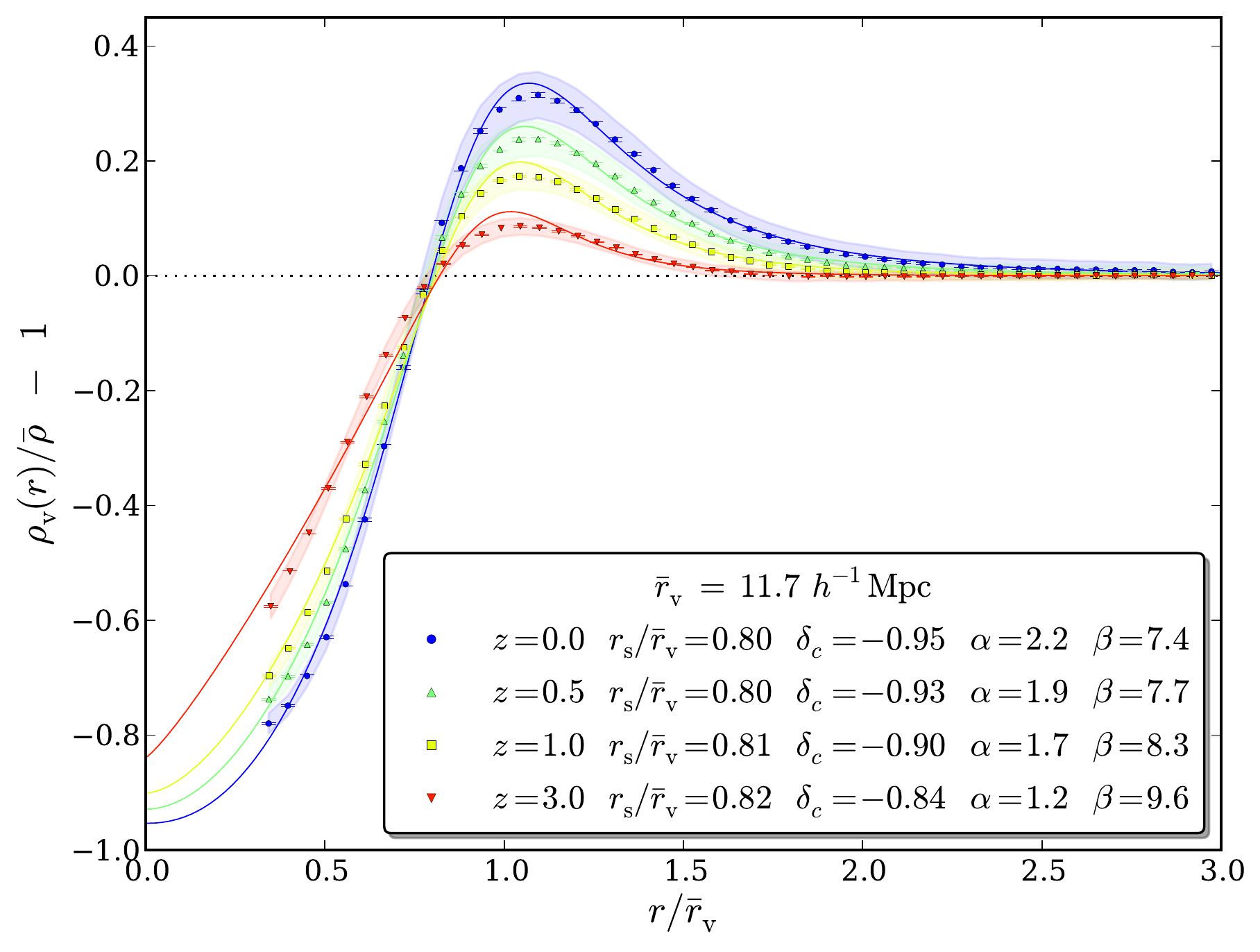}
\includegraphics[trim=0 0 0 0,clip]{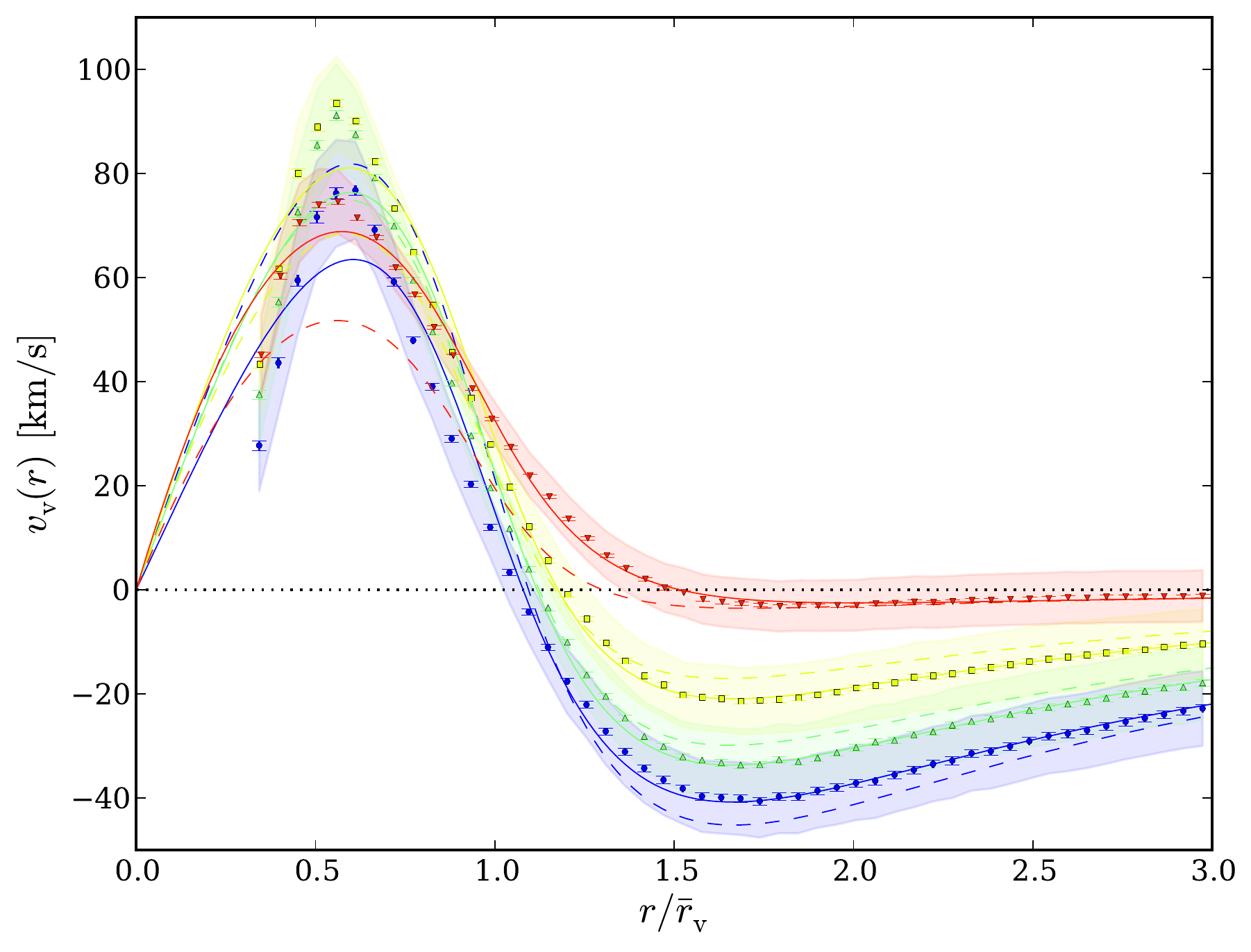}}
\caption{Same as Fig.~\ref{fig1} for voids of fixed comoving radius at different redshifts as indicated along with the best-fit parameters.}
\label{fig3}
\end{figure*}

\textit{Universality.}---Figure~\ref{fig2} depicts the best-fit parameters for each void density stack, where we plot $\delta_c$, $\alpha$, and $\beta$ against $\rs/\bar{r}_\void$. This representation reveals noticeable correlations among the parameters, which hints at a redundancy of the parameter space. In particular, $\alpha$ exhibits a linear trend with $\rs/\bar{r}_\void$, while $\beta$ follows a more complicated behavior. However, dividing the void sample into overcompensated and undercompensated voids at $\bar{r}_\void\simeq17.6\hMpc$, one can approximate $\beta$ to also follow a linear trend on either side of the vertical line indicating the compensation scale. This can be seen in Fig.~\ref{fig2}, where the solid lines result from a linear regression of the corresponding data points, with their explicit linear relations stated aside. They provide a reasonable fit with respect to the size of the error bars. For compensated voids, $\rs/\bar{r}_\void\simeq0.91$, $\alpha\simeq2$ and $\beta\simeq9.5$ attains a maximum. These values precisely satisfy Eq.~(\ref{compensated}). Even the central density $\dc$ exhibits a noticeable correlation with $\rs/\bar{r}_\void$, but following a more nonlinear behavior. We therefore do not attempt to express $\dc$ as a function of $\rs/\bar{r}_\void$ and leave it as a free parameter.

These results suggest the parametrization of Eq.~(\ref{dprofile}) to be overdetermined, and hence the number of free parameters too large for the entire sample of voids. The best-fit relations from Fig.~\ref{fig2},
\begin{gather}
\alpha(\rs)\simeq-2\left(\rs/r_\void-2\right)\;, \\
\beta(\rs)\simeq\left\{
\begin{array}{ll}
 17.5\rs/r_\void - 6.5 & \mbox{ for } \rs/r_\void < 0.91 \\
 -9.8\rs/r_\void + 18.4 & \mbox{ for } \rs/r_\void > 0.91\;,
\end{array}
\right.
\end{gather}
can be plugged back into Eq.~(\ref{dprofile}) and we can repeat the fitting procedure for the void stacks in Fig.~\ref{fig1} with the two remaining free parameters $\dc$ and $\rs$. This yields fits that are essentially indistinguishable from the original four-parameter model.

Figure~\ref{fig3} examines the redshift dependence of the density and velocity profile of voids. Here we focus on one of the previous bins with fixed comoving void radius $\bar{r}_\void=11.7\hMpc$, representing an overcompensated void. For a first-order approximation, we shall neglect the expansion or contraction of voids that can either leave or enter the bin. As apparent from the left-hand panel, the compensation walls around the void radius grow substantially, while the inner void regions are continuously emptied out, in agreement with theoretical expectations~\cite{Sheth2004}. This is consistent with the evolution of velocities as depicted in the right-hand panel of Fig.~\ref{fig3}. Because of its overcompensation, tracer particles outside the void build up higher and higher velocities towards the void center, which causes the compensation wall to grow. Inside the void the outflow also first increases at high redshift, but as the Universe accelerates its expansion due to the onset of dark energy domination ($z\sim 1$), this trend reverses and the outflow is attenuated.

Solid lines in Fig.~\ref{fig3} represent the best-fit solutions of Eqs.~(\ref{dprofile}), (\ref{vprofile}), and (\ref{intprofile}) to the data, and the corresponding parameters for the density profile are shown in the inset. The excellent agreement even at higher redshifts indicates a universal behavior of these empirical formulas. We also repeated our entire analysis based on a \textsc{wmap7} cosmology~\cite{Komatsu2011}, finding fully consistent results. However, further investigation is needed in order to explore cosmology dependence and to confirm the universality of void density profiles using higher-resolution simulations that can resolve smaller voids. Moreover, in this Letter we neglected the impact of redshift-space distortions, since void density profiles can be reconstructed in real space when statistical isotropy is assumed~\cite{Pisani2013}. Nevertheless, we find redshift-space distortions to just mildly affect the profile shapes of voids, barely degrading the quality of our fits.

\textit{Discussion.}---There are a number of cosmological applications to make use of the presented functional form of the average void density profile, for example, studies of gravitational (anti)lensing that directly probe the projected mass distribution around voids~\cite{Bolejko2013,Higuchi2013,Krause2013,Melchior2014}, which in turn may serve as a tool for constraining models of dark matter, dark energy, and modified gravity. But also considering galaxy surveys, an accurate model for the void density profile can aid in measuring the Alcock-Paczynski effect~\cite{Alcock1979,Lavaux2012,Sutter2012b} and the integrated Sachse-Wolfe effect~\cite{Granett2008,Papai2011,Ilic2013,Hernandez2013,Cai2014a,Cai2014b}, for example. This is thanks to the universal nature of Eq.~(\ref{dprofile}), which even describes voids in the distribution of galaxies remarkably well, as demonstrated in Ref.~\cite{Sutter2013a}. With that, clustering analyses based on the void model~\cite{Hamaus2014} can directly make use of the analytical form of the density profile for voids. In Ref.~\cite{Sutter2013a} it is further pointed out that the impact of tracer sparsity and bias on the definition of voids can be accounted for by simple rescaling of void sizes (see also Refs.~\cite{Arbabi-Bidgoli2002,vonBenda-Beckmann2008}). These findings corroborate other indications that cosmic voids may indeed offer new and complementary approaches to modeling fundamental aspects of the large-scale structure of our Universe.

\begin{acknowledgments}
We thank Michael Warren for providing his $N$-body simulations and Jens Jasche, Guilhem Lavaux, Florent Leclercq, Mark Neyrinck, Alice Pisani, Ravi Sheth, Joe Silk, and Rien van de Weygaert for discussions. This work made in the ILP LABEX (under reference ANR-10-LABX-63) was supported by French state funds managed by the ANR within the Investissements d'Avenir program under reference ANR-11-IDEX-0004-02. This work was also partially supported by NSF AST 09-08693 ARRA. B.D.W. is supported by a senior Excellence Chair by the Agence Nationale de Recherche (ANR-10-CEXC-004-01) and a Chaire Internationale at the Universit\'e Pierre et Marie Curie.
\end{acknowledgments}

\bibliography{ms.bib}

\begin{thebibliography}{56}
\expandafter\ifx\csname natexlab\endcsname\relax\def\natexlab#1{#1}\fi
\expandafter\ifx\csname bibnamefont\endcsname\relax
  \def\bibnamefont#1{#1}\fi
\expandafter\ifx\csname bibfnamefont\endcsname\relax
  \def\bibfnamefont#1{#1}\fi
\expandafter\ifx\csname citenamefont\endcsname\relax
  \def\citenamefont#1{#1}\fi
\expandafter\ifx\csname url\endcsname\relax
  \def\url#1{\texttt{#1}}\fi
\expandafter\ifx\csname urlprefix\endcsname\relax\def\urlprefix{URL }\fi
\providecommand{\bibinfo}[2]{#2}
\providecommand{\eprint}[2][]{\url{#2}}

\bibitem[{\citenamefont{{Granett} et~al.}(2008)\citenamefont{{Granett},
  {Neyrinck}, and {Szapudi}}}]{Granett2008}
\bibinfo{author}{\bibfnamefont{B.~R.} \bibnamefont{{Granett}}},
  \bibinfo{author}{\bibfnamefont{M.~C.} \bibnamefont{{Neyrinck}}},
  \bibnamefont{and}
  \bibinfo{author}{\bibfnamefont{I.}~\bibnamefont{{Szapudi}}},
  \bibinfo{journal}{\apjl} \textbf{\bibinfo{volume}{683}}, \bibinfo{pages}{L99}
  (\bibinfo{year}{2008}).

\bibitem[{\citenamefont{{Biswas} et~al.}(2010)\citenamefont{{Biswas},
  {Alizadeh}, and {Wandelt}}}]{Biswas2010}
\bibinfo{author}{\bibfnamefont{R.}~\bibnamefont{{Biswas}}},
  \bibinfo{author}{\bibfnamefont{E.}~\bibnamefont{{Alizadeh}}},
  \bibnamefont{and} \bibinfo{author}{\bibfnamefont{B.~D.}
  \bibnamefont{{Wandelt}}}, \bibinfo{journal}{\prd}
  \textbf{\bibinfo{volume}{82}}, \bibinfo{eid}{023002} (\bibinfo{year}{2010}).

\bibitem[{\citenamefont{{Lavaux} and {Wandelt}}(2012)}]{Lavaux2012}
\bibinfo{author}{\bibfnamefont{G.}~\bibnamefont{{Lavaux}}} \bibnamefont{and}
  \bibinfo{author}{\bibfnamefont{B.~D.} \bibnamefont{{Wandelt}}},
  \bibinfo{journal}{\apj} \textbf{\bibinfo{volume}{754}}, \bibinfo{eid}{109}
  (\bibinfo{year}{2012}).

\bibitem[{\citenamefont{{Bos} et~al.}(2012)\citenamefont{{Bos}, {van de
  Weygaert}, {Dolag}, and {Pettorino}}}]{Bos2012}
\bibinfo{author}{\bibfnamefont{E.~G.~P.} \bibnamefont{{Bos}}},
  \bibinfo{author}{\bibfnamefont{R.}~\bibnamefont{{van de Weygaert}}},
  \bibinfo{author}{\bibfnamefont{K.}~\bibnamefont{{Dolag}}}, \bibnamefont{and}
  \bibinfo{author}{\bibfnamefont{V.}~\bibnamefont{{Pettorino}}},
  \bibinfo{journal}{\mnras} \textbf{\bibinfo{volume}{426}},
  \bibinfo{pages}{440} (\bibinfo{year}{2012}).

\bibitem[{\citenamefont{{Cai} et~al.}(2014{\natexlab{a}})\citenamefont{{Cai},
  {Neyrinck}, {Szapudi}, {Cole}, and {Frenk}}}]{Cai2014a}
\bibinfo{author}{\bibfnamefont{Y.-C.} \bibnamefont{{Cai}}},
  \bibinfo{author}{\bibfnamefont{M.~C.} \bibnamefont{{Neyrinck}}},
  \bibinfo{author}{\bibfnamefont{I.}~\bibnamefont{{Szapudi}}},
  \bibinfo{author}{\bibfnamefont{S.}~\bibnamefont{{Cole}}}, \bibnamefont{and}
  \bibinfo{author}{\bibfnamefont{C.~S.} \bibnamefont{{Frenk}}},
  \bibinfo{journal}{\apj} \textbf{\bibinfo{volume}{786}}, \bibinfo{eid}{110}
  (\bibinfo{year}{2014}{\natexlab{a}}).

\bibitem[{\citenamefont{{Li}}(2011)}]{Li2011}
\bibinfo{author}{\bibfnamefont{B.}~\bibnamefont{{Li}}},
  \bibinfo{journal}{\mnras} \textbf{\bibinfo{volume}{411}},
  \bibinfo{pages}{2615} (\bibinfo{year}{2011}).

\bibitem[{\citenamefont{{Clampitt} et~al.}(2013)\citenamefont{{Clampitt},
  {Cai}, and {Li}}}]{Clampitt2013}
\bibinfo{author}{\bibfnamefont{J.}~\bibnamefont{{Clampitt}}},
  \bibinfo{author}{\bibfnamefont{Y.-C.} \bibnamefont{{Cai}}}, \bibnamefont{and}
  \bibinfo{author}{\bibfnamefont{B.}~\bibnamefont{{Li}}},
  \bibinfo{journal}{\mnras} \textbf{\bibinfo{volume}{431}},
  \bibinfo{pages}{749} (\bibinfo{year}{2013}).

\bibitem[{\citenamefont{{Spolyar} et~al.}(2013)\citenamefont{{Spolyar},
  {Sahl{\'e}n}, and {Silk}}}]{Spolyar2013}
\bibinfo{author}{\bibfnamefont{D.}~\bibnamefont{{Spolyar}}},
  \bibinfo{author}{\bibfnamefont{M.}~\bibnamefont{{Sahl{\'e}n}}},
  \bibnamefont{and} \bibinfo{author}{\bibfnamefont{J.}~\bibnamefont{{Silk}}},
  \bibinfo{journal}{\prl} \textbf{\bibinfo{volume}{111}}, \bibinfo{eid}{241103}
  (\bibinfo{year}{2013}).

\bibitem[{\citenamefont{{Cai} et~al.}(2014{\natexlab{b}})\citenamefont{{Cai},
  {Li}, {Cole}, {Frenk}, and {Neyrinck}}}]{Cai2014b}
\bibinfo{author}{\bibfnamefont{Y.-C.} \bibnamefont{{Cai}}},
  \bibinfo{author}{\bibfnamefont{B.}~\bibnamefont{{Li}}},
  \bibinfo{author}{\bibfnamefont{S.}~\bibnamefont{{Cole}}},
  \bibinfo{author}{\bibfnamefont{C.~S.} \bibnamefont{{Frenk}}},
  \bibnamefont{and}
  \bibinfo{author}{\bibfnamefont{M.}~\bibnamefont{{Neyrinck}}},
  \bibinfo{journal}{\mnras} \textbf{\bibinfo{volume}{439}},
  \bibinfo{pages}{2978} (\bibinfo{year}{2014}{\natexlab{b}}).

\bibitem[{\citenamefont{{Shandarin} et~al.}(2012)\citenamefont{{Shandarin},
  {Habib}, and {Heitmann}}}]{Shandarin2012}
\bibinfo{author}{\bibfnamefont{S.}~\bibnamefont{{Shandarin}}},
  \bibinfo{author}{\bibfnamefont{S.}~\bibnamefont{{Habib}}}, \bibnamefont{and}
  \bibinfo{author}{\bibfnamefont{K.}~\bibnamefont{{Heitmann}}},
  \bibinfo{journal}{\prd} \textbf{\bibinfo{volume}{85}}, \bibinfo{eid}{083005}
  (\bibinfo{year}{2012}).

\bibitem[{\citenamefont{{Falck} et~al.}(2012)\citenamefont{{Falck}, {Neyrinck},
  and {Szalay}}}]{Falck2012}
\bibinfo{author}{\bibfnamefont{B.~L.} \bibnamefont{{Falck}}},
  \bibinfo{author}{\bibfnamefont{M.~C.} \bibnamefont{{Neyrinck}}},
  \bibnamefont{and} \bibinfo{author}{\bibfnamefont{A.~S.}
  \bibnamefont{{Szalay}}}, \bibinfo{journal}{\apj}
  \textbf{\bibinfo{volume}{754}}, \bibinfo{eid}{126} (\bibinfo{year}{2012}).

\bibitem[{\citenamefont{{Neyrinck}}(2012)}]{Neyrinck2012}
\bibinfo{author}{\bibfnamefont{M.~C.} \bibnamefont{{Neyrinck}}},
  \bibinfo{journal}{\mnras} \textbf{\bibinfo{volume}{427}},
  \bibinfo{pages}{494} (\bibinfo{year}{2012}).

\bibitem[{\citenamefont{{Abel} et~al.}(2012)\citenamefont{{Abel}, {Hahn}, and
  {Kaehler}}}]{Abel2012}
\bibinfo{author}{\bibfnamefont{T.}~\bibnamefont{{Abel}}},
  \bibinfo{author}{\bibfnamefont{O.}~\bibnamefont{{Hahn}}}, \bibnamefont{and}
  \bibinfo{author}{\bibfnamefont{R.}~\bibnamefont{{Kaehler}}},
  \bibinfo{journal}{\mnras} \textbf{\bibinfo{volume}{427}}, \bibinfo{pages}{61}
  (\bibinfo{year}{2012}).

\bibitem[{\citenamefont{{Neyrinck} and {Yang}}(2013)}]{Neyrinck2013}
\bibinfo{author}{\bibfnamefont{M.~C.} \bibnamefont{{Neyrinck}}}
  \bibnamefont{and} \bibinfo{author}{\bibfnamefont{L.~F.}
  \bibnamefont{{Yang}}}, \bibinfo{journal}{\mnras}
  \textbf{\bibinfo{volume}{433}}, \bibinfo{pages}{1628} (\bibinfo{year}{2013}).

\bibitem[{\citenamefont{{Leclercq} et~al.}(2013)\citenamefont{{Leclercq},
  {Jasche}, {Gil-Mar{\'{\i}}n}, and {Wandelt}}}]{Leclercq2013}
\bibinfo{author}{\bibfnamefont{F.}~\bibnamefont{{Leclercq}}},
  \bibinfo{author}{\bibfnamefont{J.}~\bibnamefont{{Jasche}}},
  \bibinfo{author}{\bibfnamefont{H.}~\bibnamefont{{Gil-Mar{\'{\i}}n}}},
  \bibnamefont{and}
  \bibinfo{author}{\bibfnamefont{B.}~\bibnamefont{{Wandelt}}},
  \bibinfo{journal}{\jcap} \textbf{\bibinfo{volume}{11}}, \bibinfo{eid}{048}
  (\bibinfo{year}{2013}).

\bibitem[{\citenamefont{{Einasto}}(1965)}]{Einasto1965}
\bibinfo{author}{\bibfnamefont{J.}~\bibnamefont{{Einasto}}},
  \bibinfo{journal}{Trudy Astrofizicheskogo Instituta Alma-Ata}
  \textbf{\bibinfo{volume}{5}}, \bibinfo{pages}{87} (\bibinfo{year}{1965}).

\bibitem[{\citenamefont{{Burkert}}(1995)}]{Burkert1995}
\bibinfo{author}{\bibfnamefont{A.}~\bibnamefont{{Burkert}}},
  \bibinfo{journal}{\apjl} \textbf{\bibinfo{volume}{447}}, \bibinfo{pages}{L25}
  (\bibinfo{year}{1995}).

\bibitem[{\citenamefont{{Navarro} et~al.}(1997)\citenamefont{{Navarro},
  {Frenk}, and {White}}}]{Navarro1997}
\bibinfo{author}{\bibfnamefont{J.~F.} \bibnamefont{{Navarro}}},
  \bibinfo{author}{\bibfnamefont{C.~S.} \bibnamefont{{Frenk}}},
  \bibnamefont{and} \bibinfo{author}{\bibfnamefont{S.~D.~M.}
  \bibnamefont{{White}}}, \bibinfo{journal}{\apj}
  \textbf{\bibinfo{volume}{490}}, \bibinfo{pages}{493} (\bibinfo{year}{1997}).

\bibitem[{\citenamefont{{Moore} et~al.}(1999)\citenamefont{{Moore}, {Quinn},
  {Governato}, {Stadel}, and {Lake}}}]{Moore1999}
\bibinfo{author}{\bibfnamefont{B.}~\bibnamefont{{Moore}}},
  \bibinfo{author}{\bibfnamefont{T.}~\bibnamefont{{Quinn}}},
  \bibinfo{author}{\bibfnamefont{F.}~\bibnamefont{{Governato}}},
  \bibinfo{author}{\bibfnamefont{J.}~\bibnamefont{{Stadel}}}, \bibnamefont{and}
  \bibinfo{author}{\bibfnamefont{G.}~\bibnamefont{{Lake}}},
  \bibinfo{journal}{\mnras} \textbf{\bibinfo{volume}{310}},
  \bibinfo{pages}{1147} (\bibinfo{year}{1999}).

\bibitem[{\citenamefont{{Colberg} et~al.}(2005)\citenamefont{{Colberg},
  {Sheth}, {Diaferio}, {Gao}, and {Yoshida}}}]{Colberg2005}
\bibinfo{author}{\bibfnamefont{J.~M.} \bibnamefont{{Colberg}}},
  \bibinfo{author}{\bibfnamefont{R.~K.} \bibnamefont{{Sheth}}},
  \bibinfo{author}{\bibfnamefont{A.}~\bibnamefont{{Diaferio}}},
  \bibinfo{author}{\bibfnamefont{L.}~\bibnamefont{{Gao}}}, \bibnamefont{and}
  \bibinfo{author}{\bibfnamefont{N.}~\bibnamefont{{Yoshida}}},
  \bibinfo{journal}{\mnras} \textbf{\bibinfo{volume}{360}},
  \bibinfo{pages}{216} (\bibinfo{year}{2005}).

\bibitem[{\citenamefont{{Padilla} et~al.}(2005)\citenamefont{{Padilla},
  {Ceccarelli}, and {Lambas}}}]{Padilla2005}
\bibinfo{author}{\bibfnamefont{N.~D.} \bibnamefont{{Padilla}}},
  \bibinfo{author}{\bibfnamefont{L.}~\bibnamefont{{Ceccarelli}}},
  \bibnamefont{and} \bibinfo{author}{\bibfnamefont{D.~G.}
  \bibnamefont{{Lambas}}}, \bibinfo{journal}{\mnras}
  \textbf{\bibinfo{volume}{363}}, \bibinfo{pages}{977} (\bibinfo{year}{2005}).

\bibitem[{\citenamefont{{Ricciardelli}
  et~al.}(2013)\citenamefont{{Ricciardelli}, {Quilis}, and
  {Planelles}}}]{Ricciardelli2013}
\bibinfo{author}{\bibfnamefont{E.}~\bibnamefont{{Ricciardelli}}},
  \bibinfo{author}{\bibfnamefont{V.}~\bibnamefont{{Quilis}}}, \bibnamefont{and}
  \bibinfo{author}{\bibfnamefont{S.}~\bibnamefont{{Planelles}}},
  \bibinfo{journal}{\mnras} \textbf{\bibinfo{volume}{434}},
  \bibinfo{pages}{1192} (\bibinfo{year}{2013}).

\bibitem[{\citenamefont{{Ricciardelli}
  et~al.}(2014)\citenamefont{{Ricciardelli}, {Quilis}, and
  {Varela}}}]{Ricciardelli2014}
\bibinfo{author}{\bibfnamefont{E.}~\bibnamefont{{Ricciardelli}}},
  \bibinfo{author}{\bibfnamefont{V.}~\bibnamefont{{Quilis}}}, \bibnamefont{and}
  \bibinfo{author}{\bibfnamefont{J.}~\bibnamefont{{Varela}}},
  \bibinfo{journal}{\mnras} \textbf{\bibinfo{volume}{440}},
  \bibinfo{pages}{601} (\bibinfo{year}{2014}).

\bibitem[{\citenamefont{{Paz} et~al.}(2013)\citenamefont{{Paz}, {Lares},
  {Ceccarelli}, {Padilla}, and {Lambas}}}]{Paz2013}
\bibinfo{author}{\bibfnamefont{D.}~\bibnamefont{{Paz}}},
  \bibinfo{author}{\bibfnamefont{M.}~\bibnamefont{{Lares}}},
  \bibinfo{author}{\bibfnamefont{L.}~\bibnamefont{{Ceccarelli}}},
  \bibinfo{author}{\bibfnamefont{N.}~\bibnamefont{{Padilla}}},
  \bibnamefont{and} \bibinfo{author}{\bibfnamefont{D.~G.}
  \bibnamefont{{Lambas}}}, \bibinfo{journal}{\mnras}
  \textbf{\bibinfo{volume}{436}}, \bibinfo{pages}{3480} (\bibinfo{year}{2013}).

\bibitem[{\citenamefont{{Sutter}
  et~al.}(2014{\natexlab{a}})\citenamefont{{Sutter}, {Lavaux}, {Hamaus},
  {Wandelt}, {Weinberg}, and {Warren}}}]{Sutter2013a}
\bibinfo{author}{\bibfnamefont{P.~M.} \bibnamefont{{Sutter}}},
  \bibinfo{author}{\bibfnamefont{G.}~\bibnamefont{{Lavaux}}},
  \bibinfo{author}{\bibfnamefont{N.}~\bibnamefont{{Hamaus}}},
  \bibinfo{author}{\bibfnamefont{B.~D.} \bibnamefont{{Wandelt}}},
  \bibinfo{author}{\bibfnamefont{D.~H.} \bibnamefont{{Weinberg}}},
  \bibnamefont{and} \bibinfo{author}{\bibfnamefont{M.~S.}
  \bibnamefont{{Warren}}}, \bibinfo{journal}{\mnras}
  \textbf{\bibinfo{volume}{442}}, \bibinfo{pages}{462}
  (\bibinfo{year}{2014}{\natexlab{a}}).

\bibitem[{\citenamefont{{Sutter}
  et~al.}(2012{\natexlab{a}})\citenamefont{{Sutter}, {Lavaux}, {Wandelt}, and
  {Weinberg}}}]{Sutter2012a}
\bibinfo{author}{\bibfnamefont{P.~M.} \bibnamefont{{Sutter}}},
  \bibinfo{author}{\bibfnamefont{G.}~\bibnamefont{{Lavaux}}},
  \bibinfo{author}{\bibfnamefont{B.~D.} \bibnamefont{{Wandelt}}},
  \bibnamefont{and} \bibinfo{author}{\bibfnamefont{D.~H.}
  \bibnamefont{{Weinberg}}}, \bibinfo{journal}{\apj}
  \textbf{\bibinfo{volume}{761}}, \bibinfo{eid}{44}
  (\bibinfo{year}{2012}{\natexlab{a}}).

\bibitem[{\citenamefont{{Sutter}
  et~al.}(2014{\natexlab{b}})\citenamefont{{Sutter}, {Lavaux}, {Wandelt},
  {Weinberg}, {Warren}, and {Pisani}}}]{Sutter2013b}
\bibinfo{author}{\bibfnamefont{P.~M.} \bibnamefont{{Sutter}}},
  \bibinfo{author}{\bibfnamefont{G.}~\bibnamefont{{Lavaux}}},
  \bibinfo{author}{\bibfnamefont{B.~D.} \bibnamefont{{Wandelt}}},
  \bibinfo{author}{\bibfnamefont{D.~H.} \bibnamefont{{Weinberg}}},
  \bibinfo{author}{\bibfnamefont{M.~S.} \bibnamefont{{Warren}}},
  \bibnamefont{and} \bibinfo{author}{\bibfnamefont{A.}~\bibnamefont{{Pisani}}},
  \bibinfo{journal}{\mnras} \textbf{\bibinfo{volume}{442}},
  \bibinfo{pages}{3127} (\bibinfo{year}{2014}{\natexlab{b}}).

\bibitem[{\citenamefont{{Ceccarelli} et~al.}(2013)\citenamefont{{Ceccarelli},
  {Paz}, {Lares}, {Padilla}, and {Lambas}}}]{Ceccarelli2013}
\bibinfo{author}{\bibfnamefont{L.}~\bibnamefont{{Ceccarelli}}},
  \bibinfo{author}{\bibfnamefont{D.}~\bibnamefont{{Paz}}},
  \bibinfo{author}{\bibfnamefont{M.}~\bibnamefont{{Lares}}},
  \bibinfo{author}{\bibfnamefont{N.}~\bibnamefont{{Padilla}}},
  \bibnamefont{and} \bibinfo{author}{\bibfnamefont{D.~G.}
  \bibnamefont{{Lambas}}}, \bibinfo{journal}{\mnras}
  \textbf{\bibinfo{volume}{434}}, \bibinfo{pages}{1435} (\bibinfo{year}{2013}).

\bibitem[{\citenamefont{{Pisani} et~al.}(2013)\citenamefont{{Pisani}, {Lavaux},
  {Sutter}, and {Wandelt}}}]{Pisani2013}
\bibinfo{author}{\bibfnamefont{A.}~\bibnamefont{{Pisani}}},
  \bibinfo{author}{\bibfnamefont{G.}~\bibnamefont{{Lavaux}}},
  \bibinfo{author}{\bibfnamefont{P.~M.} \bibnamefont{{Sutter}}},
  \bibnamefont{and} \bibinfo{author}{\bibfnamefont{B.~D.}
  \bibnamefont{{Wandelt}}}, \bibinfo{journal}{ArXiv e-prints}
  (\bibinfo{year}{2013}), \eprint{1306.3052}.

\bibitem[{\citenamefont{{Nadathur} and {Hotchkiss}}(2014)}]{Nadathur2014}
\bibinfo{author}{\bibfnamefont{S.}~\bibnamefont{{Nadathur}}} \bibnamefont{and}
  \bibinfo{author}{\bibfnamefont{S.}~\bibnamefont{{Hotchkiss}}},
  \bibinfo{journal}{\mnras} \textbf{\bibinfo{volume}{440}},
  \bibinfo{pages}{1248} (\bibinfo{year}{2014}).

\bibitem[{\citenamefont{Warren}(2013)}]{Warren2013}
\bibinfo{author}{\bibfnamefont{M.~S.} \bibnamefont{Warren}}, in
  \emph{\bibinfo{booktitle}{Proceedings of SC13: International Conference for
  High Performance Computing, Networking, Storage and Analysis (SC'13)}}
  (\bibinfo{publisher}{ACM}, \bibinfo{address}{New York},
  \bibinfo{year}{2013}), pp. \bibinfo{pages}{72:1--72:12}.

\bibitem[{\citenamefont{{Ade (Planck Collaboration)}}(2013)}]{Planck2013}
\bibinfo{author}{\bibfnamefont{P.~A.~R.} \bibnamefont{{Ade (Planck
  Collaboration)}}}, \bibinfo{journal}{ArXiv e-prints}  (\bibinfo{year}{2013}),
  \eprint{1303.5076}.

\bibitem[{SDS()}]{SDSS}
\emph{\bibinfo{title}{{The Sloan Digital Sky Survey}}},
  \bibinfo{note}{\url{www.sdss.org}}.

\bibitem[{DES()}]{DES}
\emph{\bibinfo{title}{{The Dark Energy Survey}}},
  \bibinfo{note}{\url{www.darkenergysurvey.org}}.

\bibitem[{EUC()}]{EUCLID}
\emph{\bibinfo{title}{{Euclid}}},
  \bibinfo{note}{\url{http://sci.esa.int/euclid}}.

\bibitem[{\citenamefont{{Neyrinck}}(2008)}]{Neyrinck2008}
\bibinfo{author}{\bibfnamefont{M.~C.} \bibnamefont{{Neyrinck}}},
  \bibinfo{journal}{\mnras} \textbf{\bibinfo{volume}{386}},
  \bibinfo{pages}{2101} (\bibinfo{year}{2008}).

\bibitem[{\citenamefont{{Platen} et~al.}(2007)\citenamefont{{Platen}, {van de
  Weygaert}, and {Jones}}}]{Platen2007}
\bibinfo{author}{\bibfnamefont{E.}~\bibnamefont{{Platen}}},
  \bibinfo{author}{\bibfnamefont{R.}~\bibnamefont{{van de Weygaert}}},
  \bibnamefont{and} \bibinfo{author}{\bibfnamefont{B.~J.~T.}
  \bibnamefont{{Jones}}}, \bibinfo{journal}{\mnras}
  \textbf{\bibinfo{volume}{380}}, \bibinfo{pages}{551} (\bibinfo{year}{2007}).

\bibitem[{\citenamefont{{Bernardeau} et~al.}(2013)\citenamefont{{Bernardeau},
  {Pichon}, and {Codis}}}]{Bernardeau2013}
\bibinfo{author}{\bibfnamefont{F.}~\bibnamefont{{Bernardeau}}},
  \bibinfo{author}{\bibfnamefont{C.}~\bibnamefont{{Pichon}}}, \bibnamefont{and}
  \bibinfo{author}{\bibfnamefont{S.}~\bibnamefont{{Codis}}},
  \bibinfo{journal}{ArXiv e-prints}  (\bibinfo{year}{2013}),
  \eprint{1310.8134}.

\bibitem[{\citenamefont{{Bertschinger}}(1985)}]{Bertschinger1985}
\bibinfo{author}{\bibfnamefont{E.}~\bibnamefont{{Bertschinger}}},
  \bibinfo{journal}{\apjs} \textbf{\bibinfo{volume}{58}}, \bibinfo{pages}{1}
  (\bibinfo{year}{1985}).

\bibitem[{\citenamefont{{White} and {Ostriker}}(1990)}]{White1990}
\bibinfo{author}{\bibfnamefont{S.~D.~M.} \bibnamefont{{White}}}
  \bibnamefont{and} \bibinfo{author}{\bibfnamefont{J.~P.}
  \bibnamefont{{Ostriker}}}, \bibinfo{journal}{\apj}
  \textbf{\bibinfo{volume}{349}}, \bibinfo{pages}{22} (\bibinfo{year}{1990}).

\bibitem[{\citenamefont{{Hamaus} et~al.}(2014)\citenamefont{{Hamaus},
  {Wandelt}, {Sutter}, {Lavaux}, and {Warren}}}]{Hamaus2014}
\bibinfo{author}{\bibfnamefont{N.}~\bibnamefont{{Hamaus}}},
  \bibinfo{author}{\bibfnamefont{B.~D.} \bibnamefont{{Wandelt}}},
  \bibinfo{author}{\bibfnamefont{P.~M.} \bibnamefont{{Sutter}}},
  \bibinfo{author}{\bibfnamefont{G.}~\bibnamefont{{Lavaux}}}, \bibnamefont{and}
  \bibinfo{author}{\bibfnamefont{M.~S.} \bibnamefont{{Warren}}},
  \bibinfo{journal}{\prl} \textbf{\bibinfo{volume}{112}}, \bibinfo{eid}{041304}
  (\bibinfo{year}{2014}).

\bibitem[{Note1()}]{Note1}
\bibinfo{note}{\protect {G}.~Lavaux (private communication)}.

\bibitem[{\citenamefont{{Sheth} and {van de Weygaert}}(2004)}]{Sheth2004}
\bibinfo{author}{\bibfnamefont{R.~K.} \bibnamefont{{Sheth}}} \bibnamefont{and}
  \bibinfo{author}{\bibfnamefont{R.}~\bibnamefont{{van de Weygaert}}},
  \bibinfo{journal}{\mnras} \textbf{\bibinfo{volume}{350}},
  \bibinfo{pages}{517} (\bibinfo{year}{2004}).

\bibitem[{\citenamefont{{Peebles}}(1993)}]{Peebles1993}
\bibinfo{author}{\bibfnamefont{P.~J.~E.} \bibnamefont{{Peebles}}},
  \emph{\bibinfo{title}{{Principles of Physical Cosmology}}}
  (\bibinfo{year}{1993}).

\bibitem[{\citenamefont{{Komatsu} et~al.}(2011)\citenamefont{{Komatsu},
  {Smith}, {Dunkley}, {Bennett}, {Gold}, {Hinshaw}, {Jarosik}, {Larson},
  {Nolta}, {Page} et~al.}}]{Komatsu2011}
\bibinfo{author}{\bibfnamefont{E.}~\bibnamefont{{Komatsu}}},
  \bibinfo{author}{\bibfnamefont{K.~M.} \bibnamefont{{Smith}}},
  \bibinfo{author}{\bibfnamefont{J.}~\bibnamefont{{Dunkley}}},
  \bibinfo{author}{\bibfnamefont{C.~L.} \bibnamefont{{Bennett}}},
  \bibinfo{author}{\bibfnamefont{B.}~\bibnamefont{{Gold}}},
  \bibinfo{author}{\bibfnamefont{G.}~\bibnamefont{{Hinshaw}}},
  \bibinfo{author}{\bibfnamefont{N.}~\bibnamefont{{Jarosik}}},
  \bibinfo{author}{\bibfnamefont{D.}~\bibnamefont{{Larson}}},
  \bibinfo{author}{\bibfnamefont{M.~R.} \bibnamefont{{Nolta}}},
  \bibinfo{author}{\bibfnamefont{L.}~\bibnamefont{{Page}}},
  \bibnamefont{et~al.}, \bibinfo{journal}{\apj} \textbf{\bibinfo{volume}{192}},
  \bibinfo{pages}{18} (\bibinfo{year}{2011}).

\bibitem[{\citenamefont{{Bolejko} et~al.}(2013)\citenamefont{{Bolejko},
  {Clarkson}, {Maartens}, {Bacon}, {Meures}, and {Beynon}}}]{Bolejko2013}
\bibinfo{author}{\bibfnamefont{K.}~\bibnamefont{{Bolejko}}},
  \bibinfo{author}{\bibfnamefont{C.}~\bibnamefont{{Clarkson}}},
  \bibinfo{author}{\bibfnamefont{R.}~\bibnamefont{{Maartens}}},
  \bibinfo{author}{\bibfnamefont{D.}~\bibnamefont{{Bacon}}},
  \bibinfo{author}{\bibfnamefont{N.}~\bibnamefont{{Meures}}}, \bibnamefont{and}
  \bibinfo{author}{\bibfnamefont{E.}~\bibnamefont{{Beynon}}},
  \bibinfo{journal}{\prl} \textbf{\bibinfo{volume}{110}}, \bibinfo{eid}{021302}
  (\bibinfo{year}{2013}).

\bibitem[{\citenamefont{{Higuchi} et~al.}(2013)\citenamefont{{Higuchi},
  {Oguri}, and {Hamana}}}]{Higuchi2013}
\bibinfo{author}{\bibfnamefont{Y.}~\bibnamefont{{Higuchi}}},
  \bibinfo{author}{\bibfnamefont{M.}~\bibnamefont{{Oguri}}}, \bibnamefont{and}
  \bibinfo{author}{\bibfnamefont{T.}~\bibnamefont{{Hamana}}},
  \bibinfo{journal}{\mnras} \textbf{\bibinfo{volume}{432}},
  \bibinfo{pages}{1021} (\bibinfo{year}{2013}).

\bibitem[{\citenamefont{{Krause} et~al.}(2013)\citenamefont{{Krause}, {Chang},
  {Dor{\'e}}, and {Umetsu}}}]{Krause2013}
\bibinfo{author}{\bibfnamefont{E.}~\bibnamefont{{Krause}}},
  \bibinfo{author}{\bibfnamefont{T.-C.} \bibnamefont{{Chang}}},
  \bibinfo{author}{\bibfnamefont{O.}~\bibnamefont{{Dor{\'e}}}},
  \bibnamefont{and} \bibinfo{author}{\bibfnamefont{K.}~\bibnamefont{{Umetsu}}},
  \bibinfo{journal}{\apjl} \textbf{\bibinfo{volume}{762}}, \bibinfo{eid}{L20}
  (\bibinfo{year}{2013}).

\bibitem[{\citenamefont{{Melchior} et~al.}(2014)\citenamefont{{Melchior},
  {Sutter}, {Sheldon}, {Krause}, and {Wandelt}}}]{Melchior2014}
\bibinfo{author}{\bibfnamefont{P.}~\bibnamefont{{Melchior}}},
  \bibinfo{author}{\bibfnamefont{P.~M.} \bibnamefont{{Sutter}}},
  \bibinfo{author}{\bibfnamefont{E.~S.} \bibnamefont{{Sheldon}}},
  \bibinfo{author}{\bibfnamefont{E.}~\bibnamefont{{Krause}}}, \bibnamefont{and}
  \bibinfo{author}{\bibfnamefont{B.~D.} \bibnamefont{{Wandelt}}},
  \bibinfo{journal}{\mnras} \textbf{\bibinfo{volume}{440}},
  \bibinfo{pages}{2922} (\bibinfo{year}{2014}).

\bibitem[{\citenamefont{{Alcock} and {Paczynski}}(1979)}]{Alcock1979}
\bibinfo{author}{\bibfnamefont{C.}~\bibnamefont{{Alcock}}} \bibnamefont{and}
  \bibinfo{author}{\bibfnamefont{B.}~\bibnamefont{{Paczynski}}},
  \bibinfo{journal}{\nat~(London)} \textbf{\bibinfo{volume}{281}},
  \bibinfo{pages}{358} (\bibinfo{year}{1979}).

\bibitem[{\citenamefont{{Sutter}
  et~al.}(2012{\natexlab{b}})\citenamefont{{Sutter}, {Lavaux}, {Wandelt}, and
  {Weinberg}}}]{Sutter2012b}
\bibinfo{author}{\bibfnamefont{P.~M.} \bibnamefont{{Sutter}}},
  \bibinfo{author}{\bibfnamefont{G.}~\bibnamefont{{Lavaux}}},
  \bibinfo{author}{\bibfnamefont{B.~D.} \bibnamefont{{Wandelt}}},
  \bibnamefont{and} \bibinfo{author}{\bibfnamefont{D.~H.}
  \bibnamefont{{Weinberg}}}, \bibinfo{journal}{\apj}
  \textbf{\bibinfo{volume}{761}}, \bibinfo{eid}{187}
  (\bibinfo{year}{2012}{\natexlab{b}}).

\bibitem[{\citenamefont{{P{\'a}pai} et~al.}(2011)\citenamefont{{P{\'a}pai},
  {Szapudi}, and {Granett}}}]{Papai2011}
\bibinfo{author}{\bibfnamefont{P.}~\bibnamefont{{P{\'a}pai}}},
  \bibinfo{author}{\bibfnamefont{I.}~\bibnamefont{{Szapudi}}},
  \bibnamefont{and} \bibinfo{author}{\bibfnamefont{B.~R.}
  \bibnamefont{{Granett}}}, \bibinfo{journal}{\apj}
  \textbf{\bibinfo{volume}{732}}, \bibinfo{eid}{27} (\bibinfo{year}{2011}).

\bibitem[{\citenamefont{{Ili{\'c}} et~al.}(2013)\citenamefont{{Ili{\'c}},
  {Langer}, and {Douspis}}}]{Ilic2013}
\bibinfo{author}{\bibfnamefont{S.}~\bibnamefont{{Ili{\'c}}}},
  \bibinfo{author}{\bibfnamefont{M.}~\bibnamefont{{Langer}}}, \bibnamefont{and}
  \bibinfo{author}{\bibfnamefont{M.}~\bibnamefont{{Douspis}}},
  \bibinfo{journal}{\aap} \textbf{\bibinfo{volume}{556}}, \bibinfo{eid}{A51}
  (\bibinfo{year}{2013}).

\bibitem[{\citenamefont{{Hern{\'a}ndez-Monteagudo} and
  {Smith}}(2013)}]{Hernandez2013}
\bibinfo{author}{\bibfnamefont{C.}~\bibnamefont{{Hern{\'a}ndez-Monteagudo}}}
  \bibnamefont{and} \bibinfo{author}{\bibfnamefont{R.~E.}
  \bibnamefont{{Smith}}}, \bibinfo{journal}{\mnras}
  \textbf{\bibinfo{volume}{435}}, \bibinfo{pages}{1094} (\bibinfo{year}{2013}).

\bibitem[{\citenamefont{{Arbabi-Bidgoli} and
  {M{\"u}ller}}(2002)}]{Arbabi-Bidgoli2002}
\bibinfo{author}{\bibfnamefont{S.}~\bibnamefont{{Arbabi-Bidgoli}}}
  \bibnamefont{and}
  \bibinfo{author}{\bibfnamefont{V.}~\bibnamefont{{M{\"u}ller}}},
  \bibinfo{journal}{\mnras} \textbf{\bibinfo{volume}{332}},
  \bibinfo{pages}{205} (\bibinfo{year}{2002}).

\bibitem[{\citenamefont{{von Benda-Beckmann} and
  {M{\"u}ller}}(2008)}]{vonBenda-Beckmann2008}
\bibinfo{author}{\bibfnamefont{A.~M.} \bibnamefont{{von Benda-Beckmann}}}
  \bibnamefont{and}
  \bibinfo{author}{\bibfnamefont{V.}~\bibnamefont{{M{\"u}ller}}},
  \bibinfo{journal}{\mnras} \textbf{\bibinfo{volume}{384}},
  \bibinfo{pages}{1189} (\bibinfo{year}{2008}).

\end{thebibliography}
\bibliographystyle{apsrev.bst}

\end{document}